\documentclass[nofootinbib,showpacs,preprint,preprintnumbers,amsmath,amssymb]{revtex4-1}

\usepackage[T1]{fontenc} 

\usepackage{amssymb,amsmath}
\usepackage{graphicx}
\usepackage{color}
\usepackage{subfigure}
\usepackage{ulem}
\usepackage{bm}       
\usepackage{dcolumn}  

\def\bes{\begin{subequations}}
\def\ees{\end{subequations}}

\def\be{\begin{equation}}
\def\ee{\end{equation}}
\def\bea{\begin{eqnarray}}
\def\eea{\end{eqnarray}}
\def\ba{\begin{eqnarray}}
\def\ea{\end{eqnarray}}
\def\bear{\begin{array}}
\def\eear{\end{array}}
\def\p1sl{\displaystyle{\not}p_1}
\def\p2sl{\displaystyle{\not}p_2}

\def\BF{{\cal B}}

\def\enu{e^\pm \nu}
\def\munu{\mu^\pm \nu}
\def\taunu{\tau^\pm \nu}

\def\tauN{\tau^\pm N}
\def\Btolnu{B^+ \to \ell^+ \nu}
\def\Btoenu{B^\pm \to \enu}
\def\Btomunu{B^\pm \to \munu}
\def\Btotaunu{B^\pm \to \taunu}
\def\BtolN{B^\pm \to \ell^\pm N}

\def\BtotauN{B^\pm \to \tauN}
\def\tanB{\tan\beta}

\newcommand{\K}{{\widetilde {\cal K}}}
\newcommand{\G}{{\overline {\Gamma}}}
\newcommand{\bL}{{\overline {L}}}
\newcommand{\bG}{{\overline{\Gamma}}}

\usepackage{graphics}
\usepackage{graphicx}
\usepackage{dcolumn}
\usepackage{bm}
\usepackage{epsfig}

\begin{document}

\baselineskip 3.0ex
\vspace*{18pt}

\title{Decay of $B^\pm \to \tau^\pm +$ ``missing momentum'' and \\
  direct measurement of the mixing parameter $U_{\tau N}$}

\author{G.~Cveti\v{c}}
\email{gorazd.cvetic@usm.cl }
\affiliation{Department of Physics, Universidad T\'ecnica Federico
  Santa Mar\'ia, Valpara\'iso, Chile}
\author{C.~S.~Kim}
\email{cskim@yonsei.ac.kr }
\affiliation{Department of Physics and IPAP, Yonsei University, Seoul
  120-749, Korea \vspace{0.3cm}}
\author{Y.-J.~Kwon}
\email{yjkwon63@yonsei.ac.kr }
\affiliation{Department of Physics and IPAP, Yonsei University, Seoul
  120-749, Korea \vspace{0.3cm}}
\author{Y.~Yook}
\email{youngmin.yook@yonsei.ac.kr }
\affiliation{Department of Physics and IPAP, Yonsei University, Seoul
  120-749, Korea \vspace{0.3cm}}

\begin{abstract} \baselineskip 3.0ex  \vspace{1.0cm}
\noindent
We derive the decay widths for the leptonic decays of heavy charged
pseudoscalars to massive sterile neutrinos, $M^{\pm} \to \ell^{\pm} + N$, within the frameworks
involving the Standard Model and two-Higgs doublets (type II). We then
apply the result to $B^\pm \to \tau^\pm + {\rm ``missing~~ momentum"}$
of the Belle/{\sc BaBar} experimental results, in order to measure
$directly$ the relevant parameter space, including the mixing
parameter $U_{\tau N}$.
\end{abstract}

\maketitle
\flushbottom


\section{Introduction}
\label{intr}

The purely leptonic decays\footnote{Throughout this paper,
  charge-conjugate modes are implied as well unless stated otherwise.}
$\Btolnu$ have been of great interest as a probe for new physics
beyond the Standard Model (SM), because in the SM the decay rate can
be calculated very precisely and new physics effects, for instance,
charged Higgs contributions \cite{Hou} in the two-Higgs doublet
models~\cite{HHG} may appear in the tree-level contribution.  In the
SM, the decay rates are proportional to $m_\ell^2$, the square of the
corresponding charged lepton mass; therefore the $B^\pm$ decays to
$\enu$ and $\munu$ final states are highly suppressed in comparison to
$\Btotaunu$.  Since the first evidence for $\Btotaunu$ decays was
obtained by the Belle experiment~\cite{BtotaunuFirst}, its branching
fraction has been measured by Belle and {\sc BaBar}~\cite{BtotaunuBFs},
resulting in the world-average value~\cite{PDG2014}
\be \label{BtotaunuExp} {\cal B}_{\rm exp} (\Btotaunu) = (1.14 \pm
0.27)\times 10^{-4} ~.  \ee This value is consistent with the SM-based
prediction, \be \label{BtotaunuSM} {\cal B}_{\rm SM} (\Btotaunu) =
(0.758_{-0.059}^{+0.080}) \times 10^{-4} ~, \ee which is obtained by
fitting the Cabibbo-Kobayashi-Maskawa (CKM) unitarity constraints~\cite{CKMfitter}, at the level
of approximately $1.5\sigma$.  This implies that if the measurement is
improved in future $B$-factory experiments such as
Belle~II~\cite{Belle2}, the comparison can clarify whether new physics
scenarios are needed.

Heavy sterile neutral particles (also known as ``heavy neutrinos''), with
suppressed mixing with the sub-eV SM neutrinos, appear in
several new physics scenarios, such as the original seesaw
\cite{seesaw} with very heavy neutrinos, seesaw with neutrinos of mass
$0.1$-$1$ TeV \cite{1TeVNu}, or even scenarios with
neutrino mass $\sim 1$ GeV \cite{1GeVNu,nuMSM,lowscaleseesaw}.  We
will consider the reaction $\BtotauN$, where $N$ stands for the heavy
neutrino, any heavy sterile neutrino of Dirac type or Majorana type,
in interpreting the measured branching fraction in the new
physics perspective.  Even if $N$ is invisible in the detector, we can
still separate $\BtolN$ signals from $\Btolnu$ for $\ell = e$ or
$\mu$, because of the two-body nature of the decay whereby the
momentum of the charged lepton in the $B$ meson rest frame is nearly
monochromatic and depends on the mass of $N$.  The situation is
complicated for $\BtotauN$ because there is more than one neutrino
in the final state
due to the fast decay of $\tau^{\pm}$, and the decay
signature of $\BtotauN$ becomes almost indistinguishable from the
ordinary $\Btotaunu$.  Therefore, we may not exclude the possibility
that the experimentally observed signal of $\Btotaunu$ may actually
contain contributions from $\BtotauN$.

Since the measurement of $\Btotaunu$ is not only an important CKM
unitarity test of the SM, but also a very effective probe into new
physics models regarding the charged Higgs, it is important to
identify and study any unknown decay modes that can affect the
measured branching fraction of $\Btotaunu$ as much as we can.  In this
paper we analyze $\BtotauN$ both in the SM framework with a minimal
extension to include $N$ and in the frameworks involving two-Higgs
doublets (type II).

\section{Decay widths of $B^\pm \to \tau^\pm +$ ``missing momentum'' and
  determination of relevant parameters}
\label{sec:der}

Massive neutrinos may be the final state particles of the leptonic
decays of the heavy mesons (such as $B^{\pm}$), if such neutrinos mix
with the standard flavor neutrinos.
If the mixing coefficient for the heavy mass eigenstate $N$ with the
standard flavor neutrino $\nu_{\ell}$ ($\ell = e, \mu, \tau$) is
denoted as $U_{\ell N}$,\footnote{ Other notations for $U_{\ell N}$
  exist in the literature, $e.g.$ $V_{\ell 4}$ in
  \cite{Atre}; $B_{\ell N}$ in \cite{PilZPC,CDK}.}  then the standard flavor
neutrino $\nu_{\ell}$ ($\ell =e, \mu, \tau$) can be represented as \be
\nu_{\ell} = \sum_{k=1}^3 U_{\ell \nu_k} \nu_k + U_{\ell N} N \ ,
\label{mix}
\ee where $\nu_k$ ($k=1,2,3$) denote the light mass eigenstates.  In
our simplified notation above, we assumed only one additional massive
sterile neutrino $N$. The unitary extended Pontekorvo-Maki-Nakagawa-Sakata (PMNS) matrix
$U$~\cite{Pmns} is
in this case a $4 \times 4$ matrix.  However, our formulas, to be
derived in this section, will be applicable also to more extended
scenarios as well (with more than one additional massive neutrinos
$N_j$) .

The decay $B^+ \to \tau^+ N$ then proceeds via exchange of an
(off-shell) $W^+$ SM gauge boson. In addition, if the Higgs structure
involves two-Higgs doublets, the exchange of the charged Higgs $H^+$
also contributes, cf. Figs.~\ref{FigMNell} (a) and
(b), as shown in Appendix A.
Straightforward calculation, given in Appendix \ref{app1}, gives us an
expression for the decay width $\Gamma(B^+ \to \tau^+ N)$;
cf. Eq.~(\ref{Gamma2}) in conjunction with Eqs.~(\ref{rHlH}) and
(\ref{lambda}).

In the decay $B^+ \to \tau^+ \nu_{\tau}$, within the SM with $M_{\nu_{\tau}} \approx 0$ and with no
charged Higgs, only the decay mode of Fig.~\ref{FigMNell}(a)
contributes, and the expression (\ref{Gamma2}) reduces to
\be
\label{GSMnu}
\Gamma_{\rm SM}(B^+ \to \tau^+ \nu_{\tau}) =
\frac{1}{8 \pi} G_F^2 f_B^2 |V_{ub}|^2 \left( 1 -
  \frac{M_{\tau}^2}{M_B^2} \right)^2 M_B M_{\tau}^2 \ .
\ee
Here,
$M_B$ and $f_B$ are the $B^+$ meson mass and the decay constant,
respectively, $|V_{ub}|$ is the corresponding CKM matrix element, and
$G_F$ is the Fermi constant.  Using this formula, with the values
$f_B=0.1906$ GeV (i.e., the central value of $f_B=0.1906 \pm 0.0047$
GeV Ref.~\cite{PDG2014}), $M_B=5.279$ GeV and $\tau_B=1.638 \times
10^{-12}$ s \cite{PDG2014}, the SM branching ratio values
Eq.~(\ref{BtotaunuSM}) from the CKM unitarity constraints
\cite{CKMfitter} would then imply for $|V_{ub}|$ the values $|V_{ub}|
= (3.44_{-0.14}^{+ 0.18}) \times 10^{-3}$.  It is interesting that
these values are very close to the values $|V_{ub}| = (3.23 \pm 0.31)
\times 10^{-3}$ obtained from exclusive decays $B^{+} \to \pi \ell^+
\nu$ while the inclusive charmless decays give significantly different
values $|V_{ub}| = (4.41 \pm 0.22) \times 10^{-3}$ \cite{PDG2014}.

If the above decay, with $M_{\nu_{\tau}} \approx 0$, is considered
within the two-Higgs doublet model type II [2HDM(II), \cite{HHG}],
both modes Fig.~\ref{FigMNell}(a) and (b) contribute, and the
expression (\ref{Gamma2}) reduces to
\be \Gamma_{\rm 2HDM(II)}(B^+ \to
 \tau^+ \nu_{\tau}) = \frac{1}{8 \pi} G_F^2 f_B^2 |V_{ub}|^2 \left( 1 -
  \frac{M_{\tau}^2}{M_B^2} \right)^2 M_B M_{\tau}^2 r_H^2 = r_H^2
\Gamma_{\rm SM}(B^+ \to \tau^+ \nu_{\tau}) \ ,
\label{G2HDMnu}
\ee
where the factor $r_H$ is given in Appendix \ref{app1}
in Eq.~(\ref{rH}) which, in the considered
case of $B^+$ decay, is
\be
r_H = - 1 + \frac{M_B^2}{M_H^2} \tan^2 \beta  \ ,
\label{rHB}
\ee where $M_H$ is the mass of the charged Higgs $H^+$, and $\tan
\beta = v_1/v_2 = v_D/v_U$ is the ratio of the vacuum expectation
values of the two Higgs doublets (down type and up type).

Further, if we consider the decay to a massive neutrino $N$, $B^+ \to
 \tau^+ N$, the expressions (\ref{GSMnu}) and (\ref{G2HDMnu}) get
extended due to $M_N \not= 0$ and due to the mixing factor $U_{\tau
  N}$ of Eq.~(\ref{mix}).
{ However, the expressions (\ref{GSMnu}) and (\ref{G2HDMnu})
include, in addition to the channels with the first three (almost) massless
neutrinos $\nu_k$ ($k=1,2,3$), also the spurious channel with a (nonexistent)
massless fourth mass eigenstate $N'$ whose mixing coefficients are equal to that
of the (true) massive $N$, i.e., $U_{\ell N}$. This is due to the unitarity of
the $4 \times 4$ mixing matrix $U$ appearing in Eq.~(\ref{mix}) which
implies the relation
\be
\sum_{k=1}^3 |U_{\ell \nu_k}|^2 + |U_{\ell N}|^2 = 1 \ .
\label{unit}
\ee Therefore, any deviation from the values (\ref{GSMnu}) and
(\ref{G2HDMnu}) due to the existence of a massive neutrino $N$ will be
equal to $\Gamma(B^+ \to \tau^+ N) - \Gamma(B^+ \to \tau^+
N)|_{M_N=0}$, where the second term is necessary in
order to avoid double counting. We will now simply denote this
difference as $\Gamma(B^+ \to \tau^+ N)$.  According to the general
formula (\ref{Gamma2}), this is then \bes
\label{GN}
\bea
\label{SMN}
\lefteqn{\Gamma_{\rm SM}(B^+ \to  \tau^+ N) =
\frac{1}{8 \pi} G_F^2 f_B^2 |V_{ub}|^2 |U_{\tau N}|^2 M_B M_{\tau}^2
}
\nonumber\\
&& {\Bigg \{}
\lambda^{1/2} \left(1 ,\frac{M_N^2}{M_B^2}, \frac{M_{\tau}^2}{M_B^2} \right)
\frac{1}{M_B^2 M_{\tau}^2}
\left[ (M_{\tau}^2 + M_N^2)  (M_B^2 - M_N^2 - M_{\tau}^2)
+ 4 M_N^2 M_{\tau}^2 \right]
\nonumber\\ &&
- \left( 1 - \frac{M_{\tau}^2}{M_B^2} \right)^2
{\Bigg \}},
\\
\label{2HDMN}
\lefteqn{\Gamma_{\rm 2HDM(II)}(B^+ \to \tau^+ N) =
 \frac{1}{8 \pi} G_F^2 f_B^2 |V_{ub}|^2 |U_{\tau N}|^2  M_B M_{\tau}^2
}
\nonumber\\
&& {\Bigg \{}
\lambda^{1/2} \left(1 ,\frac{M_N^2}{M_B^2}, \frac{M_{\tau}^2}{M_B^2} \right)
\frac{1}{M_B^2 M_{\tau}^2}
\left[ (M_{\tau}^2 r_H^2 + M_N^2 l_H^2)  (M_B^2 - M_N^2 -
  M_{\tau}^2) - 4 r_H l_H M_N^2 M_{\tau}^2 \right]
\nonumber\\ &&
- \left( 1 - \frac{M_{\tau}^2}{M_B^2} \right)^2 r_H^2 {\Bigg \}}.
\eea
\ees
}
The function $\lambda$ appearing here is defined in Appendix
\ref{app1} in Eq.~(\ref{lambda}), and the factor $l_H$ in
Eq.~(\ref{lH}) which, in the considered case of $B^+$ decay, is
\be
l_H = 1 + \frac{M_B^2}{M_H^2} \ .
\label{lHB}
\ee
{ The expressions (\ref{GSMnu}) and (\ref{SMN}) should then
  be added in the SM case, and the expressions (\ref{G2HDMnu}) and
  (\ref{2HDMN}) should be added in the 2HDM(II) case, in order to obtain
  the full decay widths for $B^{\pm} \to \tau^{\pm} +$ ``missing
  momentum.'' }

We now summarize three possible cases of numerical interest for
the decays $B^{\pm} \to \tau^{\pm} +$ ``missing momentum,''
all in scenarios beyond the SM:

\begin{enumerate}
\item If the missing momenta are only from $\nu_\tau$ of the SM,
  and there is charged Higgs contribution in addition to the
  SM process, the decay width is determined by Eq.~(\ref{G2HDMnu}).
  Figure~\ref{plots00}~(left panel) shows the allowed regions [shown
  in dark- and pale-shaded grey (red color online) corresponding to
  $\pm 1 \sigma$ and $\pm 2 \sigma$ regions, respectively] in the
  parameter space of $M_H$ and $\tanB$ in 2HDM(II).  To determine
  the allowed regions, we compare Eq.~(\ref{BtotaunuExp}) for the
  experimental value with Eq.~(\ref{BtotaunuSM}) for the SM
  contribution in the theory value.

\item If the missing momenta are due to a sterile heavy neutrino as
  well as the SM tau neutrino, while there being no charged Higgs
  contribution, the decay width is obtained by adding
  Eqs.~(\ref{GSMnu}) and (\ref{SMN}).
  Figure~\ref{plots00}~(right panel) shows the allowed regions (with
  the same color assignment as described above) in the parameter space
  of $M_N$ and $|U_{\tau N}|$ assuming no contributions from charged
  Higgs.  Again, to determine these regions we use
  Eq.~(\ref{BtotaunuExp}) for the experimental value and
  Eq.~(\ref{BtotaunuSM}) for the SM contribution in the theory
  value. We also assume, for this figure, that the
    sterile heavy particle $N$ is invisible, and hence does not decay
    inside the detector. Note that the upper bound of the allowed
  region of $|U_{\tau N}|$ goes beyond 1, which is obviously much
  larger than the existing upper bound listed in Table~\ref{tabB2}.
  This is mainly because the central value of the current
  world average of $\BF(\Btotaunu)$ is significantly larger than the
  SM-based calculation obtained from the CKM unitarity constraints,
  Eqs.~(\ref{BtotaunuExp}) and (\ref{BtotaunuSM}).  The upper bound
    is less restrictive at low masses $M_N$, because the results for
    the process $B^{\pm} \to \tau^{\pm} +$ ``missing momentum'' are
    indistinguishable from those of SM when $M_N \to 0$.

\begin{figure}[ht]
\centering
\includegraphics[width=\textwidth]{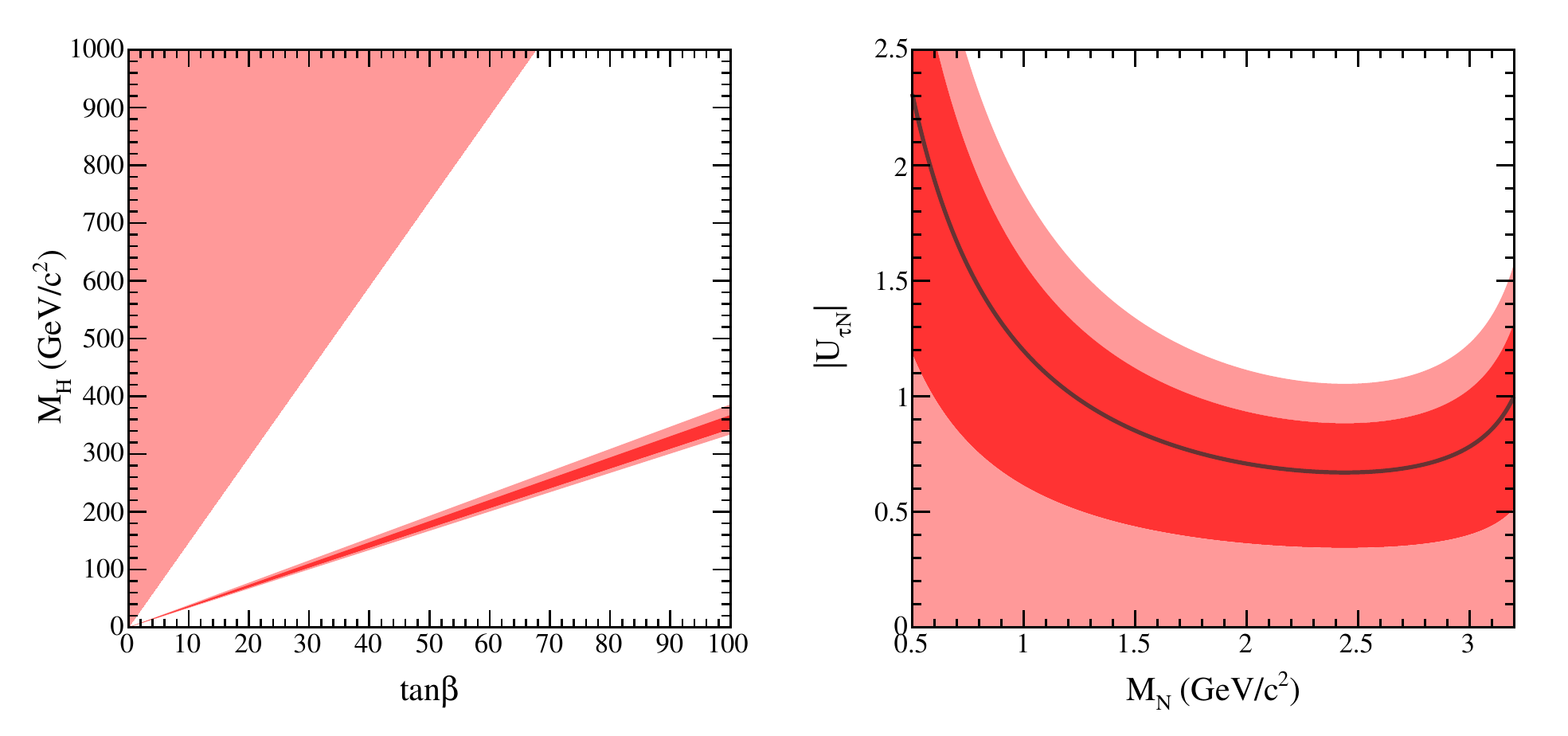}
\vspace{-0.4cm}
\caption{\baselineskip 3.0ex  The allowed regions determined from the
  measurement of $\Btotaunu$ in two cases: (left) the allowed regions in
  the parameter space of $M_{H^+}$ $vs.$ $\tanB$ in 2HDM(II) assuming
  that the missing momenta are only from $\nu_\tau$ of the SM; (right)
  The allowed regions in the parameter space of $M_N$ $vs.$ $|U_{\tau
    N}|$ assuming no contributions from charged Higgs but allowing the
  possible contributions from heavy neutrino $N$. The dark- and
  pale-shaded areas (red online) correspond to $\pm 1\sigma$ and $\pm
  2\sigma$ allowed regions, respectively.}
\label{plots00}
\end{figure}

\begin{table}[ht]
\centering
\caption{\baselineskip 3.0ex  Presently known upper bound estimates
  ($cf.$~Ref.~\cite{Atre}) for $|U_{\ell N}|^2$ ($\ell = e, \mu, \tau$)
  for  $M_N \approx 1$, $3$ GeV.}
\label{tabB2}
\begin{tabular}{c|c|c|c}
  ~~$M_N$ [GeV]~~ & ~~$|U_{e N}|^2$~~ &  ~~$|U_{\mu N}|^2$~~ & ~~$|U_{\tau N}|^2$~~  \\
  \hline\hline
  $\approx 1.0$ &  $10^{-7}$ & $10^{-7}$ & $10^{-2}$  (\cite{delphi})
  \\
  $\approx 3.0$ &  $10^{-6}$ & $10^{-4}$ & $10^{-4}$  (\cite{delphi})
\end{tabular}
\end{table}

One thing we note is that the bound on $|U_{\tau N}|$ in
Table~\ref{tabB2} has been determined $indirectly$ by
DELPHI~\cite{delphi} from the $invisible$ decay width of $Z^0$, 
i.e.  $e^+ e^- \to Z^0 \to N \bar \nu$, with $N - \nu_\tau$ mixing.
Since $\nu_\tau$ in this reaction was not explicitly identified, the
obtained bound is inclusive of other types of neutrinos.  On the other
hand, the $\Btotaunu$ mode, where $\tau$ is identified, is $directly$
related to $\tau-N$ coupling.  Therefore, any information on $|U_{\tau
  N}|$ obtained from $\Btotaunu$ is not influenced by any other types
of neutrinos, which makes a clear difference from the DELPHI result.
In this regard, even though the current bound on $|U_{\tau N}|$ from
$\BF(\Btotaunu)$ is much looser than that of DELPHI's, it will be of
great interest if the bound can be improved or evidence for
nonzero contribution from $N$ is found in the future measurements of
$\BF(\Btotaunu)$.

\item If the missing momenta are via a sterile heavy neutrino as well
  as the SM tau neutrino, and, there is also charged Higgs
  contribution, the decay width comes from adding Eqs.~(\ref{G2HDMnu})
  and (\ref{2HDMN}). We will discuss this interesting case in details later.
\end{enumerate}

Up until now, a possibly important effect of suppression due
to the survival probability was not included. Namely, if the detector
has a certain length $L$, the produced massive neutrino $N$ could
decay within the detector, producing additional particles.  The
elimination of such events from the decay width $\Gamma(M^+ \to
\ell^+ N)$ introduces a suppression factor $S_N =\exp[-t/(\tau_N
\gamma_N)]$, where $t \approx L/\beta_N$ is the time of flight of $N$
through the detector ($\beta_N$ being the velocity), and $\gamma_N=(1
- \beta_N^2)^{-1/2}$ is the time dilation (Lorentz) factor. Therefore,
the suppression factor, with which we should multiply the decay width
$\Gamma(B^+ \to \tau^+ N)$, is thus \be S_N = \exp \left[ -
  \frac{L}{\tau_N \gamma_N \beta_N} \right] \approx \exp \left[ -
  \frac{L \Gamma_N}{\gamma_N} \right] \ ,
\label{SN1}
\ee
where in the last relation we used $\beta_N \approx 1$ and
$\tau_N = 1/\Gamma_N$ [$ \equiv 1/\Gamma(N \to {\rm all})$],
in the units where $c=1=\hbar$.

Following Appendix \ref{sec:sur}, for the decays $B^{\pm} \to
\tau^{\pm} N$ and with detectors of length $L \sim 1~{\rm m}$, we
obtain the following:
\begin{itemize}
\item If $M_N \approx 1$ GeV: $S_N$ is significantly smaller than
  unity only if the value of $|U_{\tau N}|^2$ is close to its present
  (weakly restricted) upper bound ($|U_{\tau N}|^2 \sim 10^{-2}$);
  $S_N$ is close to unity otherwise. In our present numerical analysis,
  we assumed $S_N \sim 1$.
\item If $M_N \approx 3$ GeV: $S_N$ is significantly smaller than
  unity if at least one of the values of the three mixing elements
  $|U_{\ell^{'} N}|^2$ is close to its present upper bound ($|U_{e
    N}|^2 \sim 10^{-6}$, or $|U_{\mu N}|^2 \sim 10^{-4}$, or $|U_{\tau
    N}|^2 \sim 10^{-4}$); $S_N$ is close to unity otherwise. In our present numerical analysis,
  we assumed $S_N \sim 1$.
\end{itemize}

\section{Numerical Analysis and Discussions}
\label{sec:num}

In this Section, we discuss implications from possible future
measurements. Although Fig.~\ref{plots00} (right) shows that we can
set some constraints on the heavy neutrino mass and its coupling
through the measurement of $\BF(\Btotaunu)$, the existing uncertainty
is too large.  Since the Belle~II experiment is aiming to increase
the data sample by more than a factor of 50, we may expect that the
uncertainty of $\BF(\Btotaunu)$ can be reduced at least by an order of
magnitude.  Therefore, in the discussions below, we will assume that
the experimental uncertainty is improved by a factor of 10.

First we assume no contribution from charged Higgs such as in
2HDM(II).  Consequently, we consider only the diagram shown in
Fig.~\ref{FigMNell}(a) in Appendix A where both ordinary $\nu_\tau$
and heavy  neutral particle $N$
contribute.
Depending on the values of $U_{lN}$, $M_N$ and the
  detector size, the produced sterile neutrino $N$ can decay within or
  beyond the actual detector.  When $N$ 
 is produced in the decay $B^+ \to \tau^+ N$, and if it
  decays within the detector, the main signature of $N$ will be 
  $N \to l^+ \pi^-$ (if $N$ is Majorana) or 
  $N \to l^- \pi^+$  (if $N$ is Dirac 
or Majorana), 
  which will
  show experimentally as a resonance in $M(l^\pm \pi^\mp)$.  In order
  to maximize experimental sensitivity, we should analyze both the
  invisible mode and the visible decay modes for $N$, and combine the
  corresponding signal yields. Consequently, the value of $U_{lN}$
  shall be determined by the combined yields with appropriate
  corrections for efficiency and subdecay branching fractions.  While
  the invisible mode of $N$ will be analyzed by following the existing
  $B^+ \to \tau^+ \nu_\tau $ analyses of Belle and {\sc Babar}, some of the
  visible decay modes of $N$ should be explicitly analyzed in the future
  measurement.  Based on our expectation of the branching fractions of
  the visible modes (see Appendix C) and the survival suppression
  factor (see Appendix B), the correction factor for the undetected
  signal events can be obtained for each assumed values of $U_{lN}$
  and $m_N$.

\begin{figure}[ht]
\centering \includegraphics[width=\textwidth]{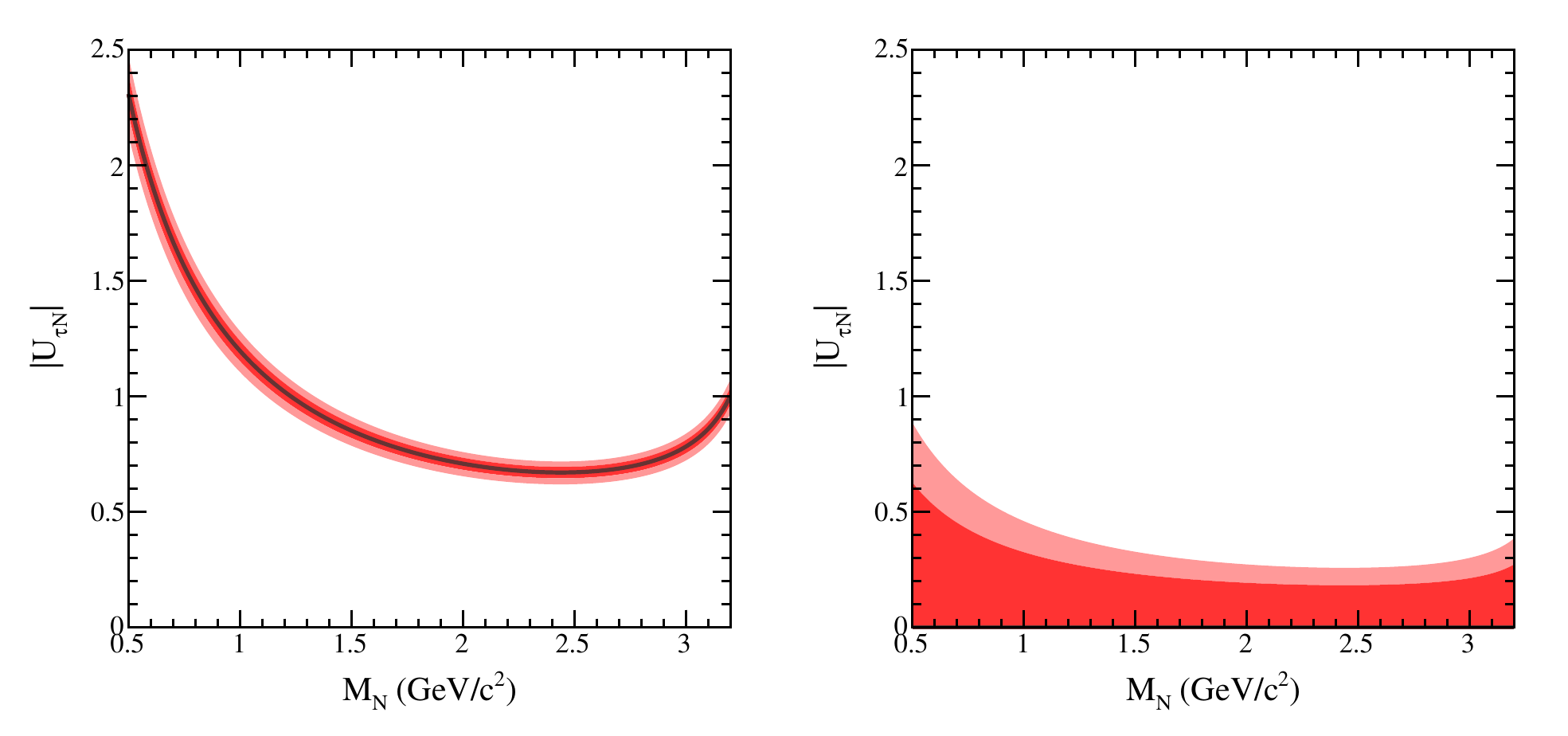}
\vspace{-0.4cm}
\caption{\baselineskip 3.0ex  The allowed regions in the parameter space
  of $M_N$ and $|U_{\tau N}|$ assuming no contributions from charged
  Higgs, where the $\pm 1\sigma$ and $\pm 2\sigma$ allowed regions are
  displayed by dark and pale shades (red online), respectively: (left)
  the central value of $\BF (\Btotaunu) $ is taken as the current world
  average Eq.~(\ref{BtotaunuExp}), while tenfold reduction of
  uncertainty is assumed; (right) the central value is taken to be the
  value predicted by the CKM unitarity constraint, also assuming tenfold
  reduction of experimental uncertainty.  In all cases, the comparison
  is made to the value determined from the CKM unitarity fitting
  Eq.~(\ref{BtotaunuSM}). }
\label{MNvsB10}
\end{figure}

Figure~\ref{MNvsB10} shows the allowed regions in the parameter space
of $M_N$ and $|U_{\tau N}|$ in this case.  The shaded area (red
online) corresponds to $\pm 1\sigma$ (dark) and $\pm 2\sigma$ (pale)
allowed regions.  For the plot on the left panel, the central value of
$\BF (\Btotaunu) $ is taken from Ref.~\cite{PDG2014}, i.e.,
Eq.~(\ref{BtotaunuExp}), but with the uncertainty reduced by a factor
of 10.  The plot on the right panel assumes that the central value is
equal to the value predicted by the CKM unitarity
constraint~\cite{CKMfitter}, i.e., Eq.~(\ref{BtotaunuSM}), again
with tenfold reduction of uncertainty (i.e., $\pm 0.027 \times
10^{-4}$).  In both cases, the $allowed$ regions are determined by
comparing the expected experimental outcome to the theory value, where
the SM contribution is taken from Eq.~(\ref{BtotaunuSM}).

From Fig.~\ref{MNvsB10} (left), it is
{evident}
that we will need an additional contribution from, e.g., $B^+ \to \tau^+
N$ if the central value of the current measurement of $\BF(\Btotaunu)$
stays the same while a substantial reduction of the measurement
uncertainty is achieved.
{Furthermore,}
comparison of
Fig.~\ref{MNvsB10} (left) with the inclusive upper bound in Table
\ref{tabB2} implies that such a scenario would require additional new
physics, e.g. 2HDM(II) charged Higgs exchange contributions.
Note that, in our numerical analysis that follows, we assume the
suppression factor $S_N = 1$.  However, with $|U_{\tau N}|^2 \sim
10^{-2}$, $\gamma_N=2$ and $L=0.1~{\rm m}$, we get $S_N \sim 1$ (for
$M_N = 1~{\rm GeV}$), but $S_N \sim 0$ (for $M_N = 3~{\rm GeV}$),
which means that the heavy sterile neutrino $N$ is
  likely to decay within the detector if $M_N \gtrsim
3~{\rm GeV}$. Therefore, it is important to consider
  visible decay modes of $N$ as well as the invisible mode.

\begin{figure}[ht]
\centering
\includegraphics[width=\textwidth]{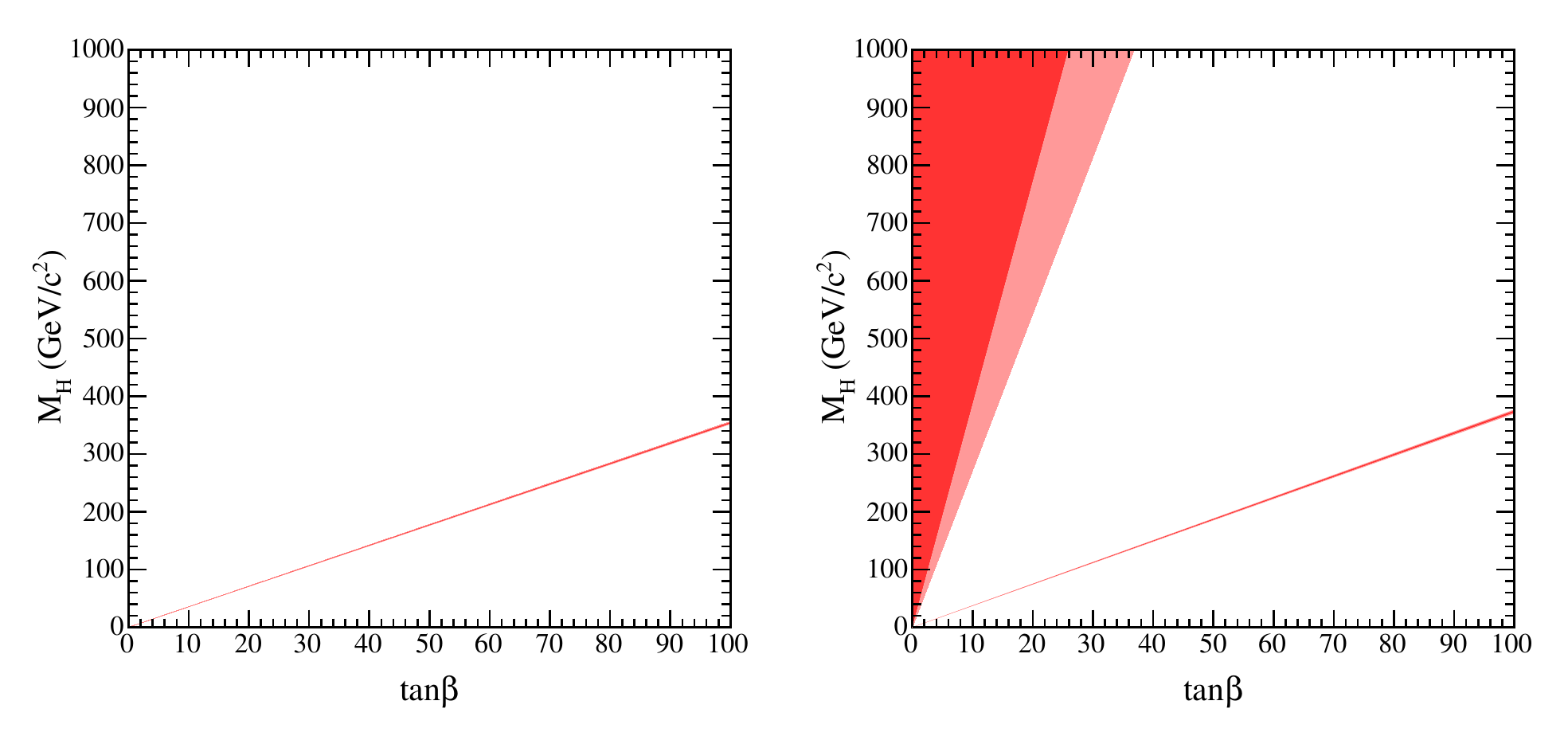}
\vspace{-0.4cm}
\caption{\baselineskip 3.0ex  The allowed regions in the parameter space
  of $M_{H^+}$ $vs.$ $\tanB$ in 2HDM (type II), assuming no
  contributions from heavy neutral particle $N$, where the $\pm 1\sigma$
  and $\pm 2\sigma$ allowed regions are displayed by dark and pale
  shades (red online), respectively: (left) the central value of $\BF
  (\Btotaunu) $ is taken from the current world average, while tenfold
  reduction of uncertainty is assumed; (right) the central value is
  taken to be the value predicted by the CKM unitarity constraint, also
  assuming tenfold reduction of experimental uncertainty.  In both
  cases, the comparison is made to the value determined from the CKM
  unitarity constraints Eq.~(\ref{BtotaunuSM}).}
\label{MHtanB0102}
\end{figure}

{On the other hand,}
if we consider the case where there is no contribution from unknown
heavy neutrino $N$, we note that the parameters $M_{H^+}$ and $\tanB$
can be much further constrained if the $\BF (\Btotaunu) $ uncertainty
is improved, e.g. by a factor of 10. Figure~\ref{MHtanB0102} shows
the allowed regions in the parameter space of $M_{H^+}$ vs $\tanB$ of
2HDM (type II) while assuming no contributions from heavy neutral
particle $N$.  The $\pm 1\sigma$ and $\pm 2\sigma$ allowed regions are
displayed by dark and pale shades (red online), respectively.  The
left panel plot uses, for the central value of $\BF (\Btotaunu) $, the
current world average and assumes tenfold reduction of uncertainty.
For the right panel plot, we consider the case of the central value
being identical with the present value predicted by the CKM unitarity
constraint and the experimental uncertainty is reduced by a factor of
10 compared to the current value $\pm 0.27 \times 10^{-4}$.  In both
cases, the comparison is made to the value determined from the CKM
unitarity constraints Eq.~(\ref{BtotaunuSM}).  Figure \ref{MHtanB0102}
(left) shows that 2HDM(II), for each $\tan \beta$, has a very narrow
interval of the corresponding allowed values of $M_H = M_H(\beta)$ if
the central experimental value of ${\cal B}(B^{\pm} \to \tau^{\pm}
\nu)$ remains approximately unchanged and the experimental uncertainty
is reduced tenfold.

\begin{figure}[ht]
\centering \includegraphics[width=0.8\textwidth]{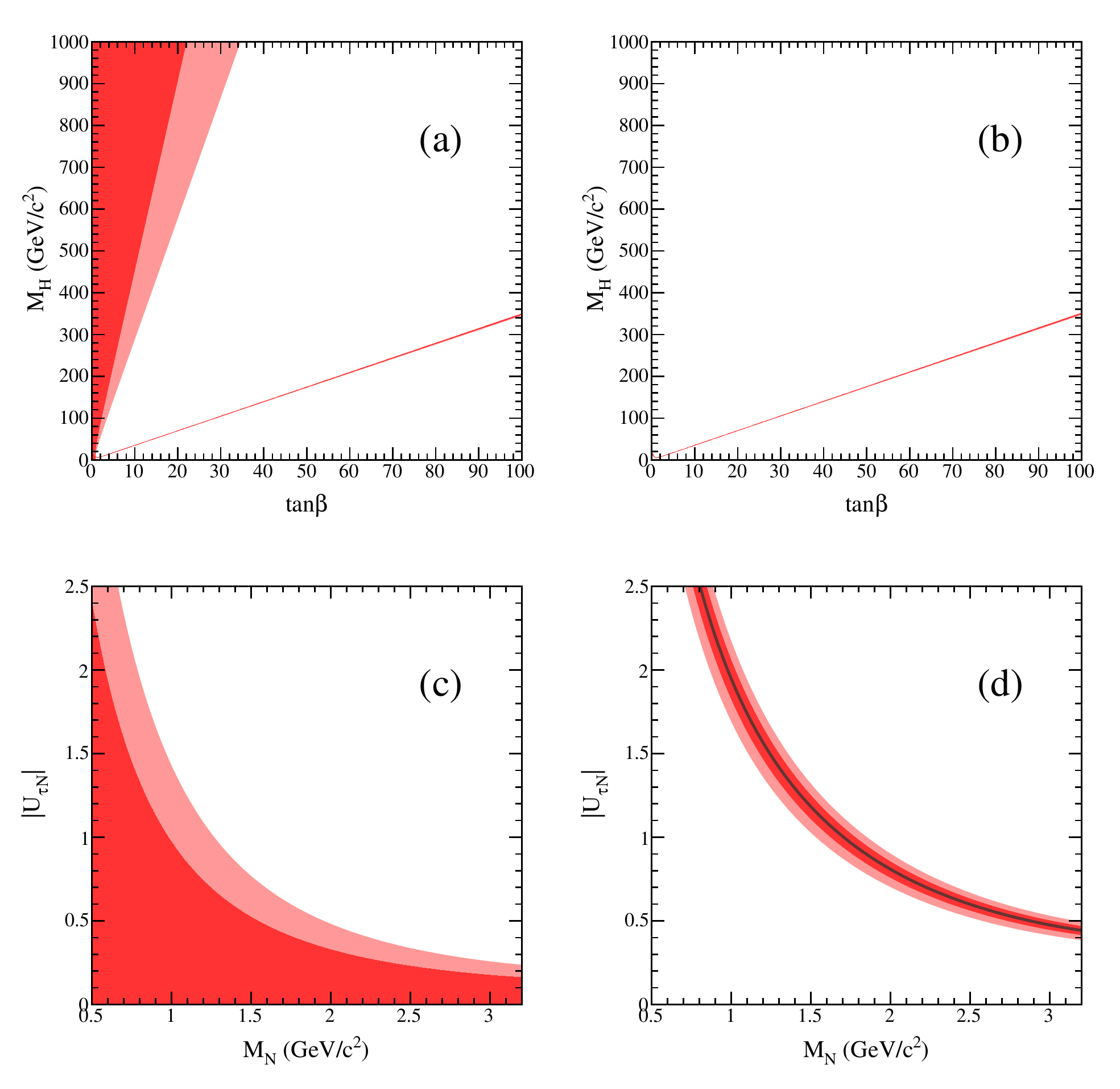}
\vspace{-0.4cm}
\caption{\baselineskip 3.0ex  The allowed regions under the assumption
  that both $H^+$ and $N$ contribute to the measured value of
  $\Btotaunu$, based on the decay rates of Eqs.~(\ref{G2HDMnu}) and
  (\ref{2HDMN}).  Top left: the allowed region in the parameter space of
  $M_{H^+}$ and $\tanB$ when $M_N = 1.0~{\rm GeV}/c^2$ and $|U_{\tau N}|
  = 0.6$.  Top right: the allowed region when $M_N = 1.0~{\rm GeV}/c^2$
  and $|U_{\tau N}| = 0.5$.  Bottom left: the allowed region in the
  parameter space of $|U_{\tau N}|$ and $M_N$ when $M_{H^+} = 200~{\rm
    GeV}/c^2$ and $\tanB = 56.5$.  Bottom right: the allowed region when
  $M_{H^+} = 200~{\rm GeV}/c^2$ and $\tanB = 55$.}
\label{MNvsB11}
\end{figure}

{Now, let's}
consider the case where both charged Higgs $H^\pm$ and heavy neutrino $N$
contribute to the measurement of $\Btotaunu$, based on the decay rates
of Eqs.~(\ref{G2HDMnu}) and (\ref{2HDMN}).
Figure~\ref{MNvsB11} shows a few exemplary cases. As in
  the cases of Fig.~\ref{MNvsB10}, we assume, in Fig.~\ref{MNvsB11},
  that both invisible and visible decays of $N$ are analyzed with
  appropriate corrections being applied to the signal yields to obtain
  the necessary branching fraction.
For each of the two plots in the top panel, we choose a point in the
parameter space of $M_N$ ${\rm vs}$ $|U_{\tau N}|$ and show the allowed
region in the parameter space of $M_{H^+}$ and $\tanB$.  For the
bottom panel, we choose points in the space of $M_{H^+}$ ${\rm vs}$ $\tanB$
and show the allowed region in $|U_{\tau N}|$ ${\rm vs}$ $M_N$.  The two
plots in the left panel correspond to the case where we choose points
within the allowed region, while we choose points outside the allowed
region for the two plots in the right panel.  For Fig.~\ref{MNvsB11}(a), 
we choose $M_N = 1.0~{\rm GeV}/c^2$ and $|U_{\tau N}| =
0.6$.  On the other hand, for Fig.~\ref{MNvsB11}(b) we choose
$M_N = 1.0~{\rm GeV}/c^2$ and $|U_{\tau N}| = 0.5$. Although it may
seem a small difference between the two cases, the resulting allowed
regions shown in the $M_{H^+}$-vs-$\tanB$ space is clearly different.
Similarly, we choose $M_{H^+} = 200~{\rm GeV}/c^2$ for both plots in
the bottom panel of Fig.~\ref{MNvsB11}, but $\tanB = 56.5$ (allowed)
for the left and $\tanB = 55$ (excluded) for the right.  Again, we see
clear difference between the two cases.

\section{Summary and Prospects}

We have seen, in Figs.~\ref{plots00}-\ref{MNvsB11}, that the
existing measurements of $\Btotaunu$ can be reinterpreted by
including the possible contributions from a heavy neutral sterile
particle $N$.  However, with the current experimental uncertainty, the
case is not clear yet.  On the other hand, with the upcoming
next-generation measurements from flavor physics facilities such as
Belle~II, the experimental uncertainty can be greatly
reduced, as was indicated in Figs. 2-4. In that case we can
set interesting constraints on parameters of 2HDM(II) and on
{heavy}
neutrino $N$ in the range $ M_N \lesssim 3~{\rm GeV}/c^2$.
For instance, if the current experimental average value of
$\BF(\Btotaunu)$ stays approximately the same but the experimental
uncertainties are greatly reduced, it can be a strong indication that
we may need a charged Higgs such as in 2HDM(II), and/or possibly a heavy
neutrino $N$.

{ In the decays $B^{\pm} \to \ell^{\pm}+ N$ when $\ell = e$ or $\mu$, the problems of double counting encountered
in Sec.~\ref{sec:der} do not appear, because the kinematics allows to distinguish such decays from those of
 $B^{\pm} \to \ell^{\pm}+ \nu_k$ ($k=1,2,3$) as mentioned in the Introduction. The relevant formulas for such cases are thus
\bes \label{GNemu}
\bea \label{SMNemu}
\Gamma_{\rm SM}(B^+ \to  \ell^+ N) &=&
\frac{1}{8 \pi} G_F^2 f_B^2 |V_{ub}|^2 |U_{\ell N}|^2
\lambda^{1/2} \left(1 ,\frac{M_N^2}{M_B^2}, \frac{M_{\ell}^2}{M_B^2} \right) \\
&& \times
\frac{1}{M_B}
\left[ (M_{\ell}^2 + M_N^2)  (M_B^2 - M_N^2 - M_{\ell}^2)
+ 4 M_N^2 M_{\ell}^2 \right] , \nonumber \,
\\
\label{2HDMNemu}
\Gamma_{\rm 2HDM(II)}(B^+ \to \ell^+ N) & = &
 \frac{1}{8 \pi} G_F^2 f_B^2 |V_{ub}|^2 |U_{\ell N}|^2
\lambda^{1/2} \left(1 ,\frac{M_N^2}{M_B^2}, \frac{M_{\ell}^2}{M_B^2} \right)
\frac{1}{M_B} \\
&&
\times \left[ (M_{\ell}^2 r_H^2 + M_N^2 l_H^2)  (M_B^2 - M_N^2 -
  M_{\ell}^2) - 4 r_H l_H M_N^2 M_{\ell}^2 \right] \ , \nonumber
\eea \ees
where $\ell = e$ or $\mu$. Although the mass of $\ell$ is in this case
small or almost zero implying the helicity suppression in the
three-generation case, the presence of a massive neutrino $M_N \sim 1$
GeV may make the decays (\ref{GNemu}) appreciable, depending certainly
on the mixing strength $|U_{\ell N}|^2$.}
In the $e^+ e^-$ $B$-factory experiments where $B$ mesons
are produced via $e^+ e^- \to \Upsilon(4S) \to B\overline{B}$ process,
the presence of heavy neutrino $N$ in the decays $B^+ \to \ell^+
N~(\ell =e, \mu)$ will be distinguishable from $B^+ \to \ell^+
\nu_\ell$ by the momentum of $\ell^+$ in the rest frame of $B^+$.

This study can be also extended to other decay modes such as $B \to
D^{(*)} \tau^+ \nu$ and $\Btolnu$ ($\ell = e$ or $\mu$).  By combining
$\Btotaunu$ and $B \to D^{(*)} \tau^+ \nu$ together, the sensitivity
of searching for $N$ can be even more enhanced.  Moreover, it is
expected that $\Btomunu $ decays can be observed in the Belle II
experiment.  This decay, unlike $\Btotaunu$, is a two-body decay mode
of $B^\pm$; hence the final-state charged lepton ($\mu^\pm$) has a
nearly monoenergetic distribution in the rest frame of the $B$ meson.
If a heavy neutrino $N$, in addition to $\nu_\mu$ of SM, also
contributes to this decay, it will change the energy distribution of
$\mu^\pm$ and its effect can be measured experimentally.  In the case of
$\Btoenu$, the SM expectation is very low, well beyond the sensitivity
of Belle II.  Nevertheless, if a heavy neutrino exists and contributes
to $\Btoenu$, it can enhance the branching fraction to be within the
experimental sensitivity of Belle II.


\acknowledgments
This work was supported in part by Fondecyt (Chile) grant 1130599, by the National Research Foundation of Korea (NRF)
grant funded by Korea government of the Ministry of Education, Science and
Technology (MEST) (No. 2011-0017430) and (No. 2011-0020333).
\\

\newpage
\appendix

\section{Derivation of the Decay Widths $\Gamma(M^\pm \to \ell^\pm N)$}
\label{app1}

\begin{figure}[htb]
\centering\includegraphics[width=100mm]{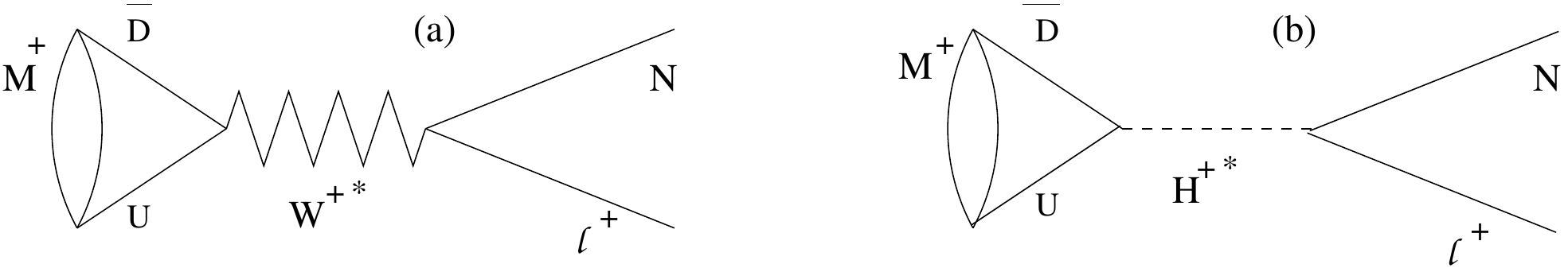}
\vspace{-0.4cm}
\caption{\baselineskip 3.0ex  The decay $M^+ \to N \ell^+$ via the
  exchange of (a) $W^+$ and (b) $H^+$.}
\label{FigMNell}
\end{figure}

\noindent In this Appendix, we derive the decay width of the process
$M^+ \to \ell^+ N$ of Figs.~\ref{FigMNell}(a) and~\ref{FigMNell} (b).
The first decay mode Fig.~\ref{FigMNell}(a),
involves an exchange of (off-shell) $W^+$.
Direct calculation gives
for the contribution ${\cal T}^{(W)}$
of the $W^+$ exchange to the reduced scattering (decay) matrix
(assuming $M_W \gg M_M$)
\be
{\cal T}^{(W)} =
U^{*}_{\ell N} V^{*}_{UD} \frac{G_F}{\sqrt{2}}
\langle 0 | {\overline D} \gamma^{\eta} (1 - \gamma_5) U | M^{+} \rangle
\left[ {\overline u}_N \gamma_{\eta} (1 - \gamma_5) v_{\ell}  \right] \ ,
\label{TW}
\ee
where ${\overline D}$ and $U$ denote the two valence quarks of the
pseudoscalar meson $M^+$, $V_{UD}$  is the corresponding CKM mixing
matrix element, and $G_F = (1/(4 \sqrt{2})) (g/M_W)^2 =
1. 166 \times 10^{-5} \ {\rm GeV}^{-2}$ .

In the model with 2HDM(II), the couplings
of the charged Higgses $H^{\pm}$ with fermions are similar  to
those of $W^{\pm}$ \cite{HHG}. The relevant parts of the Lagrange
density are
\bes
\label{Hff}
\bea
{\cal L}_{H^{\pm}qq} & = & \frac{g}{2 \sqrt{2} M_W}
\sum_{j,k} H^+ V_{U_j D_k} \left[
\cot \beta {\overline U}_j M_{U_j} (1 - \gamma_5) D_k +
\tan \beta  {\overline U}_j M_{D_k} (1 + \gamma_5) D_k
\right]
\nonumber\\
&&+ {\rm H.c.} \ ,
\label{Hqq}
\\
{\cal L}_{H^{\pm}N \ell} & = & \frac{g}{2 \sqrt{2} M_W}
\sum_{\ell} H^+ U^{*}_{\ell N} \left[
\cot \beta {\overline N} M_N (1 - \gamma_5) \ell +
\tan \beta  {\overline N} M_{\ell} (1 + \gamma_5) \ell
\right]
\nonumber\\
&&+ {\rm H.c.} \ .
\label{HNl}
\eea
\ees
The first density is for the coupling with quarks, and the second for
the the coupling with the sterile massive neutrino $N$ and the
three charged leptons $(\ell = e, \mu, \tau$). Here, $\tan \beta
= v_1/v_2 = v_D/v_U$ is the ratio of the vacuum expectation values of the two-Higgs doublets (down type and up type).

In analogy with the case of $W^+$ exchange, we obtain the contribution
${\cal T}^{(H)}$ of the $H^+$ exchange to the reduced scattering
(decay) matrix ${\cal T}$ (assuming $M_H \gg M_M$)
\bea
{\cal T}^{(H)} &=&
U^{*}_{\ell N} V^{*}_{UD} \frac{G_F}{\sqrt{2}} (-1) \frac{1}{M_H^2}
\left[
\langle 0 | {\overline D} (1 - \gamma_5) U | M^{+} \rangle
m_D \tan \beta +
\langle 0 | {\overline D} (1 + \gamma_5) U | M^{+} \rangle
m_U \cot \beta
\right]
\nonumber\\
&& \times
\left\{
\left[{\overline u}_N (1 - \gamma_5) v_{\ell}  \right] M_N \cot \beta +
\left[{\overline u}_N (1 + \gamma_5) v_{\ell}  \right] M_{\ell} \tan \beta
\right\} \ ,
\label{TH}
\eea
where $M_H$ is the mass of $H^+$.

The expressions  (\ref{TW}) and (\ref{TH}) can be further
simplified by the axial-vector current and pseudoscalar relations
\bes
\label{AVps}
\bea
\langle 0 | {\overline D} \gamma^{\eta} (1 - \gamma_5) U | M^{+} \rangle & = &
i f_M p_M^{\eta} \ ,
\label{AV}
\\
\langle 0 | {\overline D} (1 \mp \gamma_5) U | M^{+} \rangle & = &
\mp i f_M \frac{M_M^2}{m_D} \ ,
\label{ps}
\eea
\ees
where $f_M$, $M_M$  and $p_M$ are the decay constant, mass and
4-momentum of $M^+$, respectively. In Eq.~(\ref{ps}), the mass of $m_U$
was neglected in comparison with $m_D$. For example, for $B^+$ we have
$D=b$ and $U=u$.

Using the relation (\ref{AV}) in the expression (\ref{TW}), and
using the relations
\bes
\label{slashrel}
\bea
{\displaystyle{\not}p_{M}} & = &
{\displaystyle{\not}p_{N}}+{\displaystyle{\not}p_{\ell}} \ ,
\\
{\overline u}_N \displaystyle{\not}p_{N} &=& M_N {\overline u}_N \ ,
\quad
\displaystyle{\not}p_{\ell} v_{\ell} = - M_{\ell} v_{\ell} \ ,
\eea
\ees
we obtain the following form for the reduced scattering (decay)
matrix element for the $W^+$-mediated decay:
\be
{\cal T}^{(W)} =
i f_M U^{*}_{\ell N} V^{*}_{UD} \frac{G_F}{\sqrt{2}}
\left\{
+ M_N
\left[ {\overline u}_N (1 - \gamma_5) v_{\ell}  \right]
- M_{\ell}
\left[ {\overline u}_N (1 + \gamma_5) v_{\ell}  \right]
\right\}
\ .
\label{TW2}
\ee

Using the relation (\ref{ps}) in the expression (\ref{TH}),
and neglecting there the term proportional to
$m_U \cot \beta$ ($ \ll m_D \tan \beta$),
the reduced scattering (decay) matrix element for the
$H^+$-mediated decay becomes
\bea
{\cal T}^{(H)} &=&
i f_M U^{*}_{\ell N} V^{*}_{UD} \frac{G_F}{\sqrt{2}} \frac{M_M^2}{M_H^2}
\tan \beta
\nonumber\\
&& \times
\left\{
\left[{\overline u}_N (1 - \gamma_5) v_{\ell}  \right] M_N \cot \beta +
\left[{\overline u}_N (1 + \gamma_5) v_{\ell}  \right] M_{\ell} \tan \beta
\right\} \ .
\label{TH2}
\eea
Combining Eqs.~(\ref{TW2}) and (\ref{TH2}),
we obtain finally the reduced scattering (decay) matrix element ${\cal T}$
for the decay $M^+ \to N \ell^+$
\bes
\label{T}
\bea
{\cal T} & = & {\cal T}^{(W)} + {\cal T}^{(H)}
\nonumber\\
& = &
i f_M U^{*}_{\ell N} V^{*}_{UD} \frac{G_F}{\sqrt{2}}
\left\{
\left[{\overline u}_N (1 + \gamma_5) \right] M_{\ell} r_H +
\left[{\overline u}_N (1 - \gamma_5) \right] M_N l_H
\right\} \ ,
\eea
\ees
where the coefficients are
\bes
\label{rHlH}
\bea
r_H &=& - 1 + \frac{M_M^2}{M_H^2} \tan^2 \beta  \ ,
\label{rH}
\\
l_H & = & + 1 +   \frac{M_M^2}{M_H^2}  \ .
\label{lH}
\eea
\ees
The square of this matrix element (summed over the helicities of the two
final particles) then gives
\be
\langle | {\cal T} |^2 \rangle =
4 G_F^2 f_M^2 |U_{\ell N}|^2 |V_{UD}|^2
\left[ (M_{\ell}^2 r_H^2 + M_N^2 l_H^2) (p_N \cdot p_{\ell})
- 2 r_H l_H M_{\ell}^2 M_N^2 \right] \ ,
\label{Tsqr}
\ee
where we have
\be
(p_N \cdot p_{\ell}) = \frac{1}{2} (M_M^2 - M_N^2 - M_{\ell}^2) \ .
\label{pNpell}
\ee
In the rest frame of $M^+$ we then have for the decay width
\bes
\label{Gamma}
\bea
\lefteqn{
\Gamma(M^+ \to \ell^+ N)  =
\frac{1}{16 \pi M_M}
\lambda^{1/2} \left(1 ,\frac{M_N^2}{M_M^2}, \frac{M_{\ell}^2}{M_M^2} \right)
\langle | {\cal T} |^2 \rangle
}
\label{Gamma1}
\\
& = & \frac{1}{8 \pi} G_F^2 f_M^2 |U_{\ell N}|^2 |V_{UD}|^2
\lambda^{1/2} \left(1 ,\frac{M_N^2}{M_M^2}, \frac{M_{\ell}^2}{M_M^2} \right)
\nonumber\\
&& \times
\frac{1}{M_M}
\left[ (M_{\ell}^2 r_H^2 + M_N^2 l_H^2)  (M_M^2 - M_N^2 - M_{\ell}^2)
- 4 r_H l_H M_N^2 M_{\ell}^2 \right] \ ,
\label{Gamma2}
\eea
\ees
where we used the notation
\be
\lambda(y_1,y_2,y_3) =  y_1^2 + y_2^2 + y_3^2 - 2 y_1 y_2 - 2 y_2 y_3 - 2 y_3 y_1 \ .
\label{lambda}
\ee
The expression (\ref{Gamma2}), in conjunction with the expressions
(\ref{rHlH}) and (\ref{lambda}), is the
explicit expression for the decay width $\Gamma(M^+ \to N \ell^+)$
in the rest frame of $M^+$, in terms of the masses $M_M$, $M_N$, $M_{\ell}$,
$M_{H}$ and $\tan \beta \equiv v_U/v_D$.

It is straightforward to check that in the case of $M_N=0$ and
$|U_{\ell N}|=1$, the obtained formula (\ref{Gamma2}) reduces to the
formula obtained in Ref.~\cite{Hou}. Further, if $M_H \to \infty$
(i.e., no charge Higgs interchange), the formula (\ref{Gamma2})
reduces to Eq.~(2.5) of Ref.~\cite{CDK}.

\section{Survival Suppression Factor}
\label{sec:sur}


We first mention that the factor of ``nonsurvival'' probability
$P_N = 1 - S_N$ has been  discussed and investigated in the literature
for the processes where the intermediate on-shell particle (such as $N$)
is assumed to decay within the detector,
cf.~Refs.~\cite{Gronau,CDK,scatt3,CKZ1,Kimcomm,CKZ2,CDKZ,CERN-SPS}.
Here in this Section we follow the notations and results of
Ref.~\cite{CKZ2}.
The total decay width $\Gamma_N$ of sterile massive neutrino $N$ appearing in
Eq.~(\ref{SN1}) can be expressed as
\be
 \Gamma_N = \G(M_N) \K \ ,
\label{GN1}
\ee
where
\begin{equation}
 \G(M_{N}) \equiv \frac{G_F^2 M_{N}^5}{96 \pi^3} \ ,
\label{barG}
\ee
and the factor $\K$ is proportional to
the heavy-light mixing factors $|U_{\ell^{'} N}|^2$ [where
$U_{\ell^{'} N}$ appear in the relation (\ref{mix})]
\begin{equation}
\K(M_{N}) \equiv \K = {\cal N}_{e N} \; |U_{e N}|^2 + {\cal N}_{\mu N} \; |U_{\mu N}|^2 + {\cal N}_{\tau N} \; |U_{\tau N}|^2  \ .
\label{calK}
\end{equation}
Here, the coefficients
${\cal N}_{\ell^{'} N}(M_N) \equiv {\cal N}_{\ell^{'} N}$
($\ell^{'} = e, \mu, \tau$) are
the effective mixing coefficients. We have
${\cal N}_{\ell^{'} N} \sim 10^0$-$10^1$.
They are
functions of the mass $M_N$ and were numerically evaluated in Ref.~\cite{CKZ2}
for the Majorana neutrino $N$, on the basis of formulas of Ref.~\cite{HKS}.
The numerical results for the Dirac $N$ were
included in Ref.~\cite{CDKZ}.

In the ranges of $M_N$ typical for the $B^{\pm} \to N \ell^{\pm}$ decays,
i.e., for $M_N \approx 1$-$4$ GeV, we have ${\cal N}_{e N} \approx
{\cal N}_{\mu N} \approx 8$ and ${\cal N}_{\tau N} \approx 3$,
and therefore
\be
\K \approx 8  (|U_{e N}|^2 + |U_{\mu N}|^2) + 3  |U_{\tau N}|^2 \quad
(B, B_c \; {\rm decays}) \ .
\label{KapprB}
\ee
The estimate (\ref{KapprB}) is valid for Majorana $N$. For Dirac
$N$ it is somewhat lower, but the difference can be ignored
at the level of precision of the estimate, for
$1 \ {\rm GeV} \leq M_N \leq 3 \ {\rm GeV}$.

We refer to Ref.~\cite{CKZ2} (and references therein) for more details
on these results.
Furthermore, approximate values of the presently known upper bounds for the
squares of the mixing elements, in this range of
masses $M_N$, are given in Table \ref{tabB2} (cf.~Ref.~\cite{Atre}).

The survival factor (\ref{SN1}) can be rewritten as
\be
S_N = \exp \left(- \frac{L}{L_N} \right) =
\exp \left( - \frac{L}{{\bL}_N} \K \right) \ ,
\label{SN2}
\ee
where $L_N$ is the decay length, and ${\bL}_N$ is the
canonical decay length (canonical in the sense that it is independent of
the mixing parameters $U_{\ell^{'} N}$)
\bes
\label{LNbLN}
\bea
L_N^{-1} &=& {\bL}_N^{-1} \K \ ,
\label{LN}
\\
{\bL}_N^{-1} & = & \frac{\G(M_N)}{\gamma_N} \ ,
\label{bLN}
\eea
\ees
where $\G(M_N)$ is given in Eq.~(\ref{barG}). The inverse canonical
decay length $\bL_N^{-1}$, for $\gamma_N =2$, is given in Fig.~\ref{bLNfig}
as a function of $M_N$.
Specifically, we obtain $\bL_N^{-1} \approx 10^2 \ {\rm m}^{-1}$,
$3 \times 10^4 \ {\rm m}^{-1}$, for $M_N=1$ GeV, $3$ GeV, respectively.
Combining this result with the results (\ref{KapprB}) and Table \ref{tabB2},
we obtain for the effective inverse decay length $L_N^{-1}$ [appearing
in the survival factor $S_N$ of Eq.~(\ref{SN2})] the following
estimates, in units of $m^{-1}$:
\bes
\label{estLN}
\bea
L_N^{-1} (M_N \approx 1{\rm GeV}) & \approx &
0.8 \times 10^3 |U_{e N_j}|^2 + 0.8 \times 10^3 |U_{\mu N_j}|^2
\quad (+  2 \times 10^2 |U_{\tau N_j}|^2)
\nonumber\\
&\lesssim & 10^{-4} + 10^{-4}\quad (+ 10^{0}) \ ,
\label{estLND}
\\
L_N^{-1}(M_N\approx 3{\rm GeV}) & \approx &
3 \times 10^5 |U_{e N_j}|^2 + 3 \times 10^5 |U_{\mu N_j}|^2
\quad (+  1 \times 10^5 |U_{\tau N_j}|^2)
\nonumber\\
&\lesssim & 10^{0} + 10^{0}\quad (+ 10^{0}) \ .
\label{estPNB}
\eea
\ees
\begin{figure}[ht]
\centering\includegraphics[width=100mm]{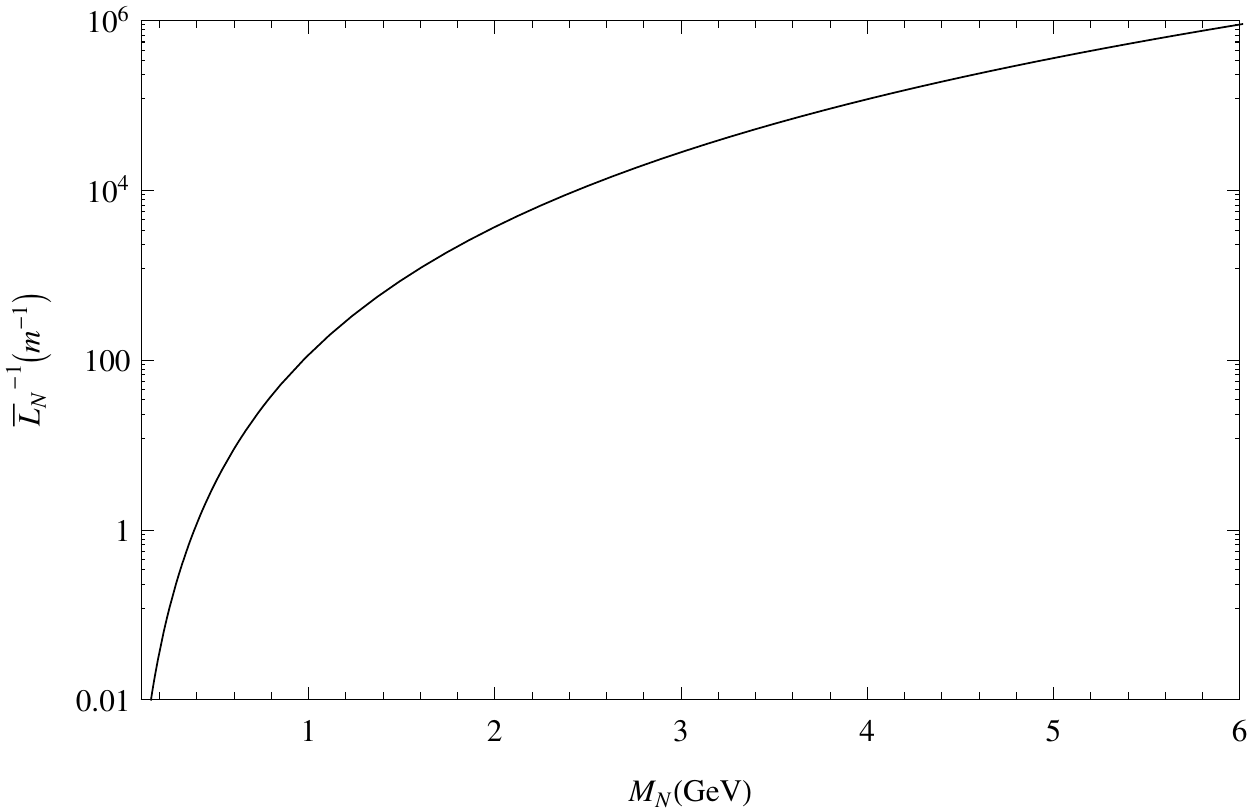}
\vspace{-0.4cm}
\caption{\baselineskip 3.0ex  The inverse canonical decay length
$\bL_N^{-1} \equiv \bG(M_{N})/\gamma_{N}$, in units of $m^{-1}$,
as a function of the neutrino mass $M_N$, for the choice
$\gamma_{N}$ [$\equiv (1 - \beta_N^2)^{-1/2}$] $=2$.}
\label{bLNfig}
\end{figure}

\section{Branching ratios for semihadronic decays of neutrino}
\label{sec:br}

Here we summarize some of the formulas for the decay widths
and branching ratios for
the decays of a heavy neutrino $N$ into hadrons \cite{Atre,HKS}.
Comparatively appreciable channels are decays into light mesons (with mass $M_H < 1$ GeV), which can be presudoscalar (P) or vector (V) mesons:
\bes
\label{GNmes}
\bea
2 \Gamma(N \to  \ell^- P^+) & = &
|B_{\ell N}|^2 \frac{G_F^2}{8 \pi} M_N^3 f_P^2 |V_P|^2 F_P(x_{\ell}, x_P) \
\label{GNMesPch}
\\
\Gamma(N \to  \nu_{\ell} P^0) & = &
|B_{\ell N}|^2 \frac{G_F^2}{64 \pi} M_N^3 f_P^2 (1 - x_P^2)^2 \
\label{GNMesP0}
\\
2 \Gamma(N \to  \ell^- V^+) & = &
|B_{\ell N}|^2 \frac{G_F^2}{8 \pi} M_N^3 f_V^2 |V_V|^2 F_V(x_{\ell}, x_V) \
\label{GNMesVch}
\\
\Gamma(N \to  \nu_{\ell} V^0) & = &
|B_{\ell N}|^2 \frac{G_F^2}{2 \pi} M_N^3 f_V^2 \kappa_V^2 (1 - x_V^2)^2
(1 + 2 x_V^2) .
\label{GNMesV0}
\eea
\ees
Here, $\ell = e, \mu, \tau$ stands generically for a charged lepton.
The charged meson channels above were multiplied by a factor $2$, because
if $N$ is Majorana neutrino both decays $N \to  \ell^- M^{+}$ and
$N \to  \ell^+ M^{-}$ contribute ($M=P, V$) equally.
The factors $f_P$ and $f_V$ are the decay constants,
and $V_P$ and $V_V$ are the CKM matrix elements involving the valence quarks
of the corresponding mesons.

In Eqs.~(\ref{GNmes}), the following notation is used:
$x_{Y} \equiv M_Y/M_N$ ($Y = \ell,  P, V$), and the
expressions $F_P$ and $F_V$ are
\bes
\label{kinex}
\bea
F_P(x,y) & = & \lambda^{1/2}(1,x^2,y^2) \left[(1 + x^2)(1 + x^2-y^2) - 4 x^2
\right] \
\label{FP}
\\
F_V(x,y) & = & \lambda^{1/2}(1,x^2,y^2) \left[(1 - x^2)^2 + (1 + x^2) y^2 - 2 y^4 \right] \ ,
\label{FV}
\eea
\ees
where
\be
\label{lam}
\lambda(y_1,y_2,y_3) =  y_1^2 + y_2^2 + y_3^2 - 2 y_1 y_2 - 2 y_2 y_3 - 2 y_3 y_1 \ .
\ee
The (light) mesons for which formulas (\ref{GNmes}) can be applied are:
$P^{\pm} = \pi^{\pm}, K^{\pm}$; $P^0 = \pi^0, K^0, {\bar K}^0, \eta$;
$V^{\pm} = \rho^{\pm}, K^{* \pm}$; $V^0= \rho^0, \omega, K^{*0}, {\bar K}^{*0}$.
If $M_N > 1$ GeV, the neutrino $N$ can decay into heavier mesons,
and the decay widths for such decays can be calculated by using duality,
as decay widths into quark pairs; nonetheless,
such decay modes are in general suppressed by kinematics
and are not given here.
The corresponding branching ratios are obtained by dividing the
above decay widths by the total decay width $\Gamma_N$ of $N$.

\end{document}